\documentstyle[epsfig,psfig]{mn2e}

\title{The {\it AKARI} Deep Field South: {\em {\it Spitzer}} 24 and 70$\mu$m Observations, Catalogs and Counts}
\author[D.L. Clements et al.]
       {D.L. Clements$^1$, G. Bendo$^1$, C. Pearson$^{2,3,4}$, Sophia A. Khan$^{1,5}$, S. Matsuura$^6$, \newauthor M. Shirahata$^6$\\
        $^1$Imperial College, London, Blackett Lab, Prince Consort Road, London SW7 2AZ, UK\\
        $^2$Rutherford Appleton Laboratory, Chilton, Didcot, Oxfordshire OX11 0QX, UK\\
        $^3$Department of Physics, University of Lethbridge, 4401 University Drive, Lethbridge, Alberta T1J 1B1, Canada\\
        $^4$Department of Physics \& Astronomy, The Open University, U.K. \\
        $^5$Shanghai Key Lab for Astrophysics, Shanghai Normal University, Shanghai 200234, China\\
        $^6$Institute of Space and Astronautical Science, JAXA, 3-1-1 Yoshinodai, Sagamihara, Kanagawa 229-8510, Japan\\
}
\date{}

\pagerange{\pageref{firstpage}--\pageref{lastpage}}
\pubyear{}

\begin{document}

\maketitle

\label{firstpage}

\begin{abstract}
The {\it AKARI} Deep Field South (ADF-S) is a $\sim$12 sq. deg. region near the South Ecliptic Pole that has been observed with deep scans in the far-infrared by the {\it AKARI} satellite. As such it is becoming one of the key extragalactic survey fields. We here present complementary observations of the ADF-S conducted by the {\it {\it Spitzer}} Space Telescope at wavelengths of 24 and 70 $\mu$m. We extract source catalogs at each of these wavelengths reaching depths of $\sim$ 0.2mJy at 24$\mu$m and $\sim$ 20mJy at 70$\mu$m. We also apply an K-to-24$\mu$m colour criterion to select objects with galaxy-like colours in the 24$\mu$m survey. Completeness corrections as a function of flux density are derived for both catalogs by injecting artificial sources of known flux density into the maps, and we find that our surbveys are 50\% complete at 0.26mJy and 24mJy at 24 and 70$\mu$m respectively. We can thus produce number counts as a function of flux density for the ADF-S at 24 and 70$\mu$m. These are combined with existing literature counts and compared to four different number count models derived from galaxy evolution models. One complicating factor for the ADF-S counts is the presence of a foreground galaxy cluster at z=0.04 in the field. We examine the ranges of flux densities to which this cluster might make a contribution to the counts and find hints that the 24$\mu$m luminosity function of the cluster galaxies might be enhanced above that of field galaxies. Full catalogs for these ADF-S {\it {\it Spitzer}} surveys at 24 and 70$\mu$m are made available as part of this paper.
\end{abstract}

\begin{keywords}
infrared:galaxies --- galaxies:evolution---galaxies:starburst
\end{keywords}

\section{Introduction}

The Japanese {\it AKARI} mission (Murakami et al., 2007) has carried out large legacy observations in high visibility areas at the North and South Ecliptic Poles. At the South Ecliptic Pole (SEP),  {\it AKARI} has made a deep survey at far-infrared wavelengths with its Far-Infrared Surveyor instrument (FIS, Kawada et al., 2007)  at wavelengths of 65, 90, 140 \& 160$\mu$m to 5$\sigma$ sensitivities of 46, 15, 180 \& 600mJy respectively (Shirahata et al., 2009, PI: S. Matsuura). The  {\it AKARI} SEP survey, hereafter referred to as the  {\it AKARI} Deep Field South (ADF-S) covers approximately 12 square degrees  centred at R.A.=4$^{h}$44$^{m}$00$^{s}$, Dec=-53deg 20$^{\prime}$00.$^{\prime \prime}$0 (J2000) (Matsuura et al., 2010), a region where the cirrus brightness is spectacularly low $< B_{100} > \sim 0.2$ MJy/sr (Schlegel et al., 1998). The main science drivers for the survey are the analysis of clustering and large scale structure using the large contiguous area, and fluctuation analysis of the background. In addition to the far-infrared data, shallow mid-infrared data also exists over the entire area, along with deep optical $R$-band imaging to 25th magnitude. The entire ADF-S has also been covered by the Balloon-borne Large-Aperture Submillimeter Telescope (BLAST,  Pascale et al. (2008) at 250, 350 \& 500$\mu$m. Other ancillary data on smaller scales include a central area of approximately 1 square degree covered by the  {\it AKARI} Infra-Red Camera (IRC, Onaka et al., 2007) at wavelengths of 3, 4, 7, 11, 15 \&24 $\mu$m, optical UBVI imaging, submillimetre observations with the Atacama Pathfinder Experiment (APEX) using the Large Bolometer Camera (LABOCA), the Atacama Submilimetre Telescope Experiment (ASTE) using the Astronomical Thermal Emission Camera (AzTEC), and 20cm Australia Telescope Compact Array (ATCA) radio observations (White et al., 2009).

{\it {\it Spitzer}} Space Telescope (Werner et al., 2004) observations of this field, covering almost the entire 12 square degrees, were also conducted. These observations were carried out with the Multiband Imaging Photometer for {\it Spitzer} (MIPS, Rieke et al., 2004) instrument at wavelengths of 24 and 70$\mu$m. Observations in the 160$\mu$m MIPS channel were not collected since the {\it AKARI} data at 140$\mu$m is at comparable depth. We here present our processing of the {\it {\it Spitzer}} ADF-S data and the catalogs extracted from our reduced images. The catalogs are made available for use by the astronomical community to maximize the scientific return from this {\it {\it Spitzer}} data and from the ADF-S in general.

As well as producing the {\it Spitzer} images and catalogs for the ADF-S, we also extract number counts at 24 and 70$\mu$m for this field and use them to set extra constraints on models of the galaxy population at these wavelengths. Previous work on counts at 24$\mu$m includes both deep counts from small areas (Papovich et al., 2004; Le Floc'h et al. 2009) and shallower counts from the largest {\it Spitzer} survey, the Spitzer Wide-Area Infrared Extragalactic Survey (SWIRE, Shupe et al., 2008) while Bethermin et al. (2010) present a compendium of counts at 24$\mu$m over a wide range of fluxes, alongside 70 and 160 $\mu$m counts. The observations we present here are reasonably well matched to the SWIRE survey counts in terms of depth, going somewhat deeper but covering only one quarter of the SWIRE area. Previous 70$\mu$m counts have concentrated more on small area fairly deep surveys, including COSMOS (Frayer et al., 2009), and GOODS-N (Frayer et al., 2006a). The largest area survey so far fully published is the FLS field (Frayer et al., 2006b) which covers 4 square degrees to depths comparable to our ADF-S {\it Spitzer} data. The SWIRE survey counts at 70$\mu$m have yet to be fully analyzed and published, though preliminary results for high completeness (S/N $>$10) SWIRE 70$\mu$m sources are presented by Frayer et al. (2009). Our work thus provides useful additional constraints for both the 24 and 70$\mu$m counts largely at bright and intermediate flux densities.

One problem that arises in the ADF-S field which has an impact on counts studies is the presence of a foreground rich galaxy cluster in the field, DC 0428-53 (also known as Abell S0463)
(Dressler, 1980; Dressler \& Schectman, 1988) at a redshift of $\sim$ 0.04. Galaxies in this cluster will tend to enhance the brighter counts, and we observe this effect. Any large area survey will be subject to the sample variance induced by large scale structures, so this is not a unique problem. Nevertheless, the presence of this cluster affects our interpretation of the brightest counts. It also yields the opportunity to study known cluster members in the infrared - something we plan to do in future publications.

The remainder of this paper is organized as follows. In Section 2 we describe the {\it Spitzer} observations and the data processing steps used to produce the reduced maps. In Section 3 we describe source extraction at 24$\mu$m and the production of the 24$\mu$m counts. Section 4 does the same at 70$\mu$m. Section 5 describes the source catalogs produced using these data, while our results are discussed and the counts compared to a variety of models in Section 6. In Section 7 we draw our conclusions. Throughout this paper we assume the concordance cosmology with H$_0 = 72 kms^{-1} Mpc^{-1}, \Omega_m = 0.3$ and $\Omega_{\Lambda} = 0.7$.

\section{{\it Spitzer} Observations and Data Processing}

The {\it {\it Spitzer}} observations were performed using the MIPS scan map
mode under programme ID 50581 PI: Mark Devlin and consisted of a total of 91.2h of observing time.  
The instrument can simultaneously acquire data in all three
MIPS bands (24, 70, and 160~$\mu$m).  Since only the 24 and 70~$\mu$m
data were needed for comparison to the {\it AKARI} data, the
instrument was used in a cryogen-conserving mode whereby the telescope
warmed up slightly to a point where the 160~$\mu$m data are not
usable.  The observations consist of a total of 34 Astronomical
Observation Requests (AORs).  32 of the AORs were scan map
observations with 9 scan legs that were 1.5$^\circ$ long and that were
offset from each other by 160~arcsec.  The other 2 AORs were scan map
observations with 5 scan legs that were 0.5$^\circ$ long and that
were offset from each other by 160~arcsec.  All scan maps were
performed at the medium scan rate, and all scans were approximately
parallel to the short axis of the SEP field.  The observed field is
approximately $6.5^\circ \times 2.2^\circ$ and contains 11.8 square
degrees of data at 24~$\mu$m and 11.5 square degrees of data at
70~$\mu$m. Execution of these AORs was spread over the period from
2008 September 24 to 2008 October 1. Typical total exposure time per map pixel after these
observations were completed was 164s at 24$\mu$m and 73.7s at 70$\mu$m.

The 24 and 70 micron
images were created from raw data frames using the MIPS Data Analysis
Tools (DAT) version 3.10 (Gordon et al. 2005) along with additional
processing steps.  The data processing for the
two wave bands differs significantly, so the processing for each band
is described in separate paragraphs below.

The individual 24~$\mu$m frames were first processed through a droop
correction (removing an excess signal in each pixel that is
proportional to the signal in the entire array) and were corrected for
non-linearity in the ramps.  The dark current was then subtracted.
Next, scan-mirror-position dependent flats were created from the data
in each AOR and were applied to the
data.  Detector pixels that had measured signals of 2500 DN s$^{-1}$
in any frame were masked out in the following three frames so as to
avoid having latent images appear in the data.  Next, a
scan-mirror-position independent flat was created from the data in
each AOR and were applied to the data.  Following this, we subtracted
variations in the background related to the position of individual
frames in the scan legs, which removes zodiacal light emission and
variations in the background related to the scan mirror position as
well as any other cyclical background fluctuations.  Next, a robust
statistical analysis was applied in which the values of cospatial
pixels from different frames were compared to each other and
statistical outliers (e.g. probable cosmic rays) were masked out.
After this, a final mosaic was made with pixel sizes of 4~arcsec$^2$,
and the data were multiplied by the flux density calibration
factor $(4.54\pm0.18) \times 10^{-2}$ MJy sr$^{-1}$ [MIPS instrumental
unit]$^{-1}$ (Engelbracht et al., 2007). 

At the beginning of the 70~$\mu$m data processing, ramps were fit to
the reads to derive slopes.  In this step, readout jumps and cosmic
ray hits were also removed, and an electronic nonlinearity correction
was applied.  Next, the stim flash frames (frames of data in which a
calibration light source was flashed at the detectors) were used as
responsivity corrections.  After this, the dark current was subtracted
from the data, and an illumination correction was applied.  Following
this, short term variations in the signal (often referred to as
drift) were removed, and additional periodic variations in the
background related to the stim flash cycle were subtracted; this also
subtracted the background from the data.  Next, a robust statistical
analysis was applied to cospatial pixels from different frames in
which statistical outliers (which could be pixels affected by cosmic
rays) were masked out.  Once this was done, final mosaics were made
using pixel sizes of 4~arcsec$^2$.  Finally, the data were multiplied
by the flux density calibration factor $702 \pm 35$ MJy sr${^-1}$ [MIPS
instrumental unit]$^{-1}$ (Gordon et al., 2007).

The final 24$\mu$m image together with a coverage map is shown in Fig. 1, while Fig. 2 shows the final 70$\mu$m image and coverage map. A magnified section of the image at each wavelength is shown in Fig. 3 where it is easier to see the full range of fluxes of the objects detected in the survey.

\begin{figure*}
\label{24map}
\epsfig{file=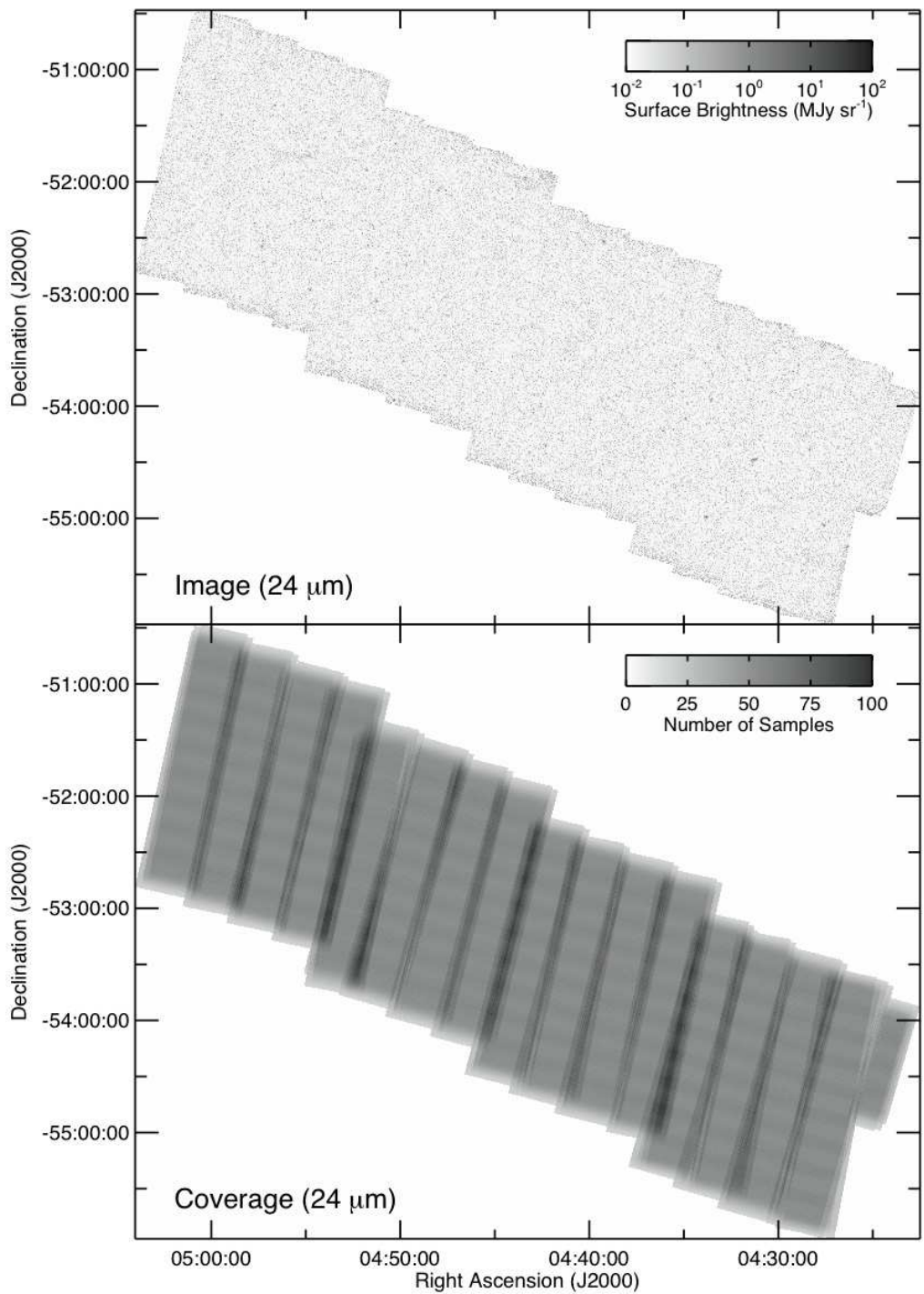,height=20cm,angle=0}
\caption{The ADF-S Field at 24$\mu$m. Top: 24$\mu$m image. Bottom: Coverage map at 24$\mu$m. Darker colours indicate higher flux  density or greater coverage respectively.}
\end{figure*}

\begin{figure*}
\label{70map}
\epsfig{file=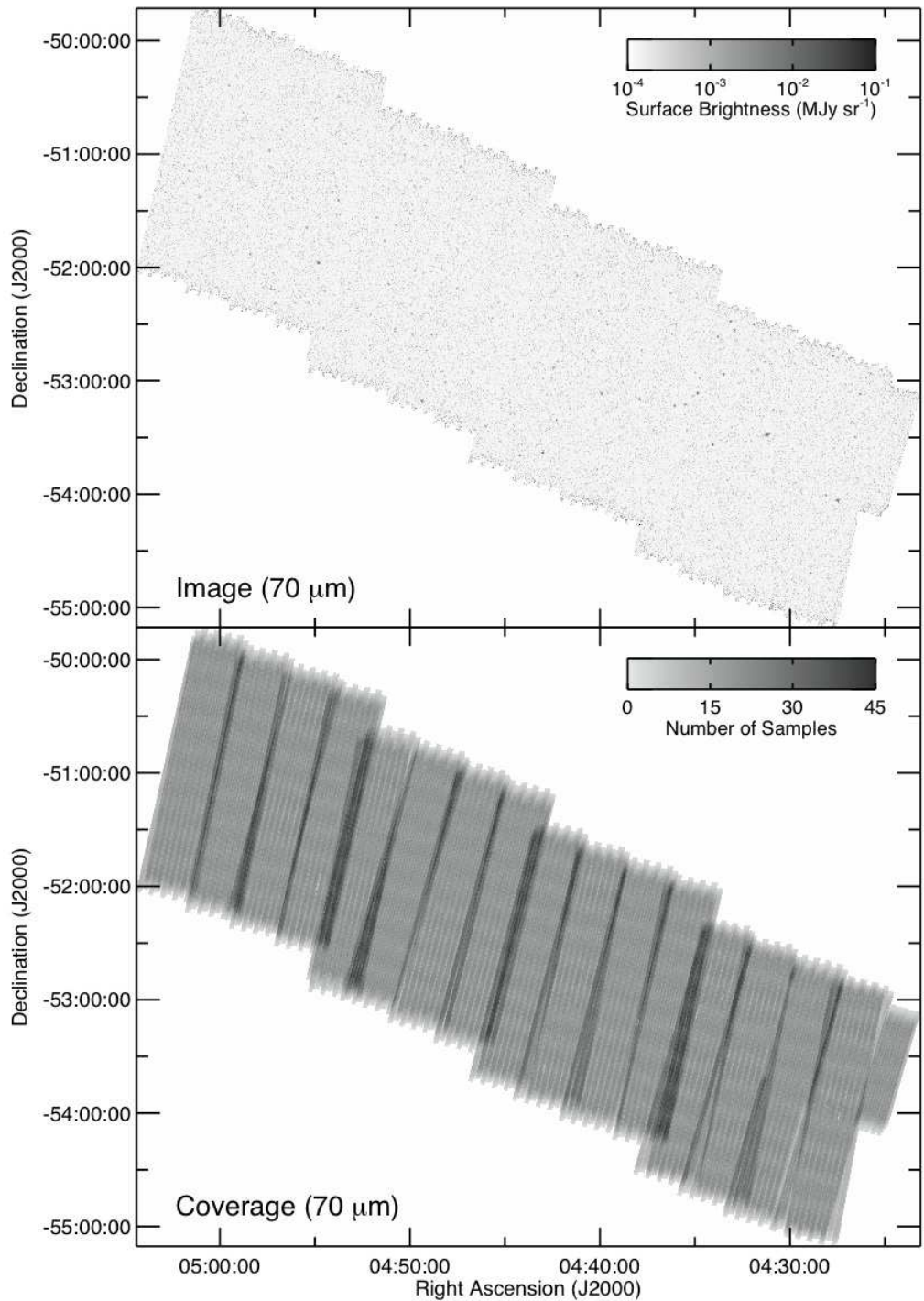,height=20cm,angle=0}
\caption{The ADF-S Field at 70$\mu$m. Top: 70$\mu$m image. Bottom: Coverage map of 70$\mu$m. Darker colours indicate higher flux  density or greater coverage respectively.}
\end{figure*}

\begin{figure*}
\label{blownmap}
\epsfig{file=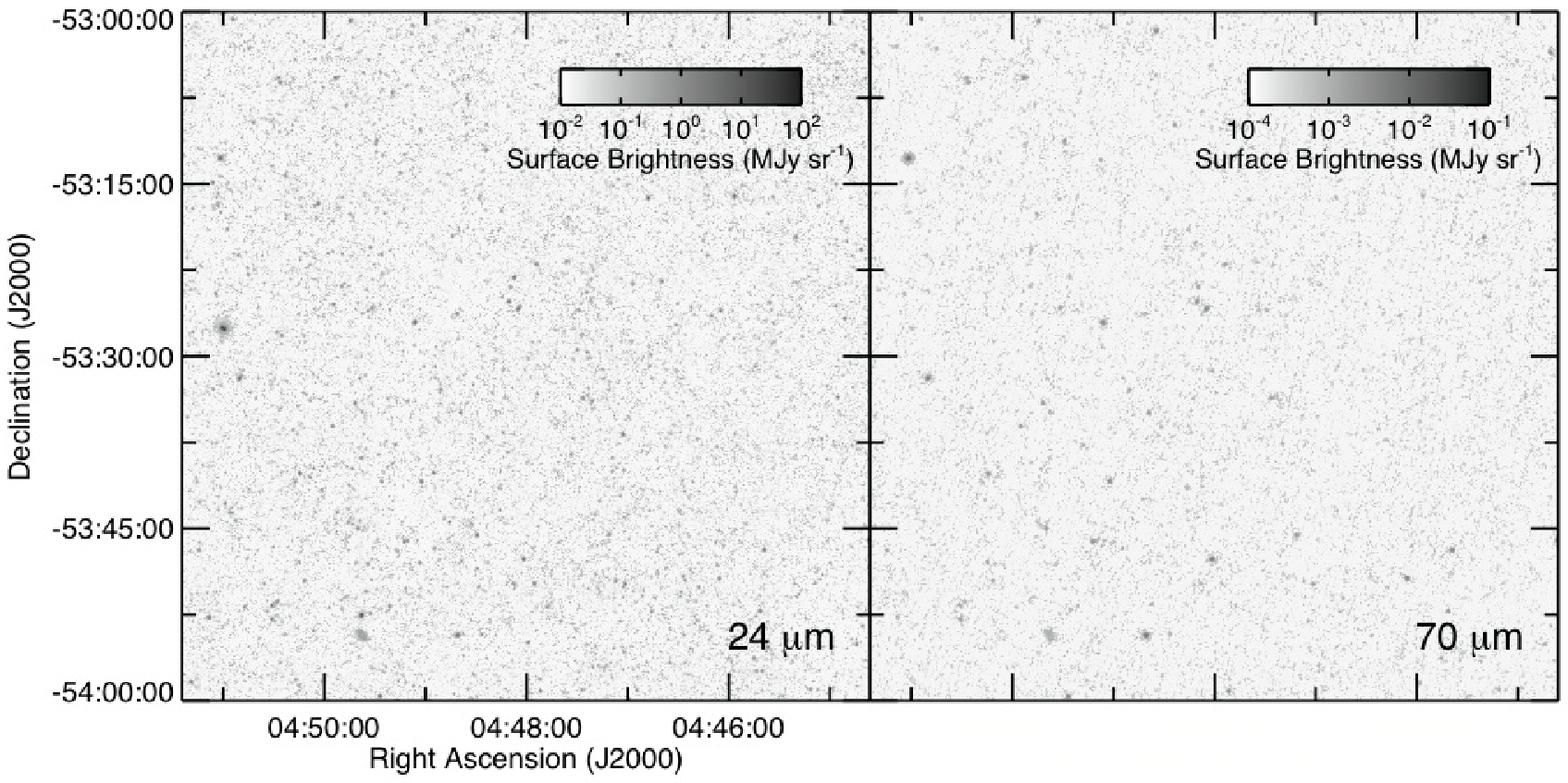,height=10cm,angle=0}
\caption{Enlarged section of ADF-S Field at 24$\mu$m (left) and 70$\mu$m (right) allowing  more of the fainter sources to be seen.}
\end{figure*}

\section{24 Micron Source Extraction and Counts}

\subsection{Source Detection and Extraction}

For the detection and extraction of sources at 24$\mu$m, we follow the recipe given in Shupe et al. (2008) since our data is broadly similar to the 24$\mu$m SWIRE survey data dealt with in that work. Source extraction was performed using SExtractor (Bertin \& Arnouts, 1996). We use a local background calculated over a region of size 128 $\times$ 128 pixels. Sources were detected using a 2$\sigma$ threshold with a minimum five connected pixels above this threshold. Aperture photometry using a 5.25 arcsec aperture and a Kron flux (Kron 1980) were extracted for each source. The flux assigned to a source was the aperture flux unless the measured size of the source was $>$100 pixels and the SExtractor stellarity probability was less than 80\%. If these were both true the Kron flux was used instead of the aperture flux, as in Shupe et al. (2008). We only extract sources from regions covered by at least ten MIPS frames. This results in the detection of 41503 sources in a region of 11.26 sq. deg. 

We follow the SWIRE team in applying a correction factor of 1.15 to match the MIPS team calibrations to bright stars (Surace et al., 2005). We also adopt their aperture correction factor of 1.55 to account for light outside the point source 5.25 arcsec aperture, a factor of 0.961 to correct the photometry to an assumed $F_{\nu} \sim \nu^{-1}$ source spectrum (Stansberry et al., 2007) and an additional correction factor of 1.018  to account for changes in the 24$\mu$m MIPS calibration factor (Surace et al., 2005; Shupe et al., 2008; MIPS Data Handbook, 2006)).

\subsection{Star-Galaxy Separation}

At 24$\mu$m flux densities brighter than a few mJy stars are a potentially serious contaminant for investigations of extragalactic number counts (Shupe et al., 2008). To separate stars from galaxies, we cross match the sources in our 24$\mu$m catalog to sources in the 2MASS K band catalog (Skrutskie et al., 2006) using a 3 arcsecond search radius. Of a total of 41503 24$\mu$m sources extracted in areas covered more than 10 times by MIPS, we obtain K band matches to 2206. The resulting K-[24] colour magnitude diagram is shown in Figure 4, where we adopt [24] as a 24$\mu$m magnitude with a zero point of 7.43 Jy in colour corrected units. We use a colour cut of K-[24]$<2$, as shown in Figure 4 and similar to Shupe et al. (2008), to exclude stars. This leaves us with 40556 sources with galaxy-like colours, and 947 objects with star-like colours. The latter are excluded from the main number count analysis. Differential counts for the sources identified as stars are shown in the Appendix.

\begin{figure}
\label{stargal}
\epsfig{file=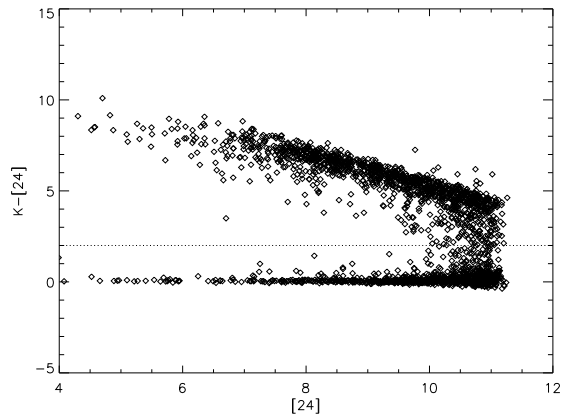,height=8cm,angle=90}
\caption{Colour-Magnitude Diagram of 2MASS K band and [24] for the ADF-S field.
The dashed line shows the colour cut used to separate out the stellar contribution to the number counts (below the line).}
\end{figure}

\subsection{Completeness Correction and Reliability Test}

The completeness of our 24$\mu$m catalog is assessed by adding artificial sources of known flux to the map and then determining the fraction of these sources recovered by the extraction process. A thousand artificial sources are added at random positions at each of nine different flux density values using an empirical noiseless point spread function (PSF) image. This empirical PSF was created using archival Spitzer data of 3C 273,
3C 279, and BL Lac.  Data from individual AORs for these three
galaxies were processed into images using the MIPS DAT and then
combined so as to filter out any extended, asymmetric emission.  See
Young et al. (2009) for additional details. Sources were then extracted from the map as described above, and we determined what fraction of the input sources were recovered. The resulting completeness correction as a function of flux is shown in Fig. 5. We use linear interpolation from this data to calculate completeness corrections for the differential number counts. We find that our catalog is complete at the 80\% and 50\% levels at fluxes of 0.32 and 0.26 mJy respectively.
We also use this data to check the flux density correction factors we have adopted from the SWIRE team ie. the 1.15 correction to bright stars and the 1.55 aperture correction factor for a total correction factor of 1.78 before colour corrections are included. We find that this flux density correction factor works well at all flux densities except those below $\sim$0.3 mJy where our completeness is very poor.

To assess the reliability of our source extraction we do the following. Firstly we subtract the mean per-pixel value of the 24$\mu$m map from the map. This produces a map with mean value zero. We then multiply this map by -1 and add back the mean we have subtracted. This produces a map with identical noise and statistical properties to the real map, but with all the real sources made negative. Sources detected by SExtractor in this map are thus false, resulting from the noise properties of the map, and the number of such sources detected can be used to assess the reliability of the source catalog. We then process the resulting false source catalog in an identical way to the real data, with the exception of star galaxy separation since there is no reason to expect the false sources to be associated with real objects on the sky. We find that our reliability is close to 100\% for the full flux range used for counts, with our faintest and most unreliable flux bin, at 0.285mJy, still having a reliability of 96\%.

\begin{figure}
\label{comp24}
\epsfig{file=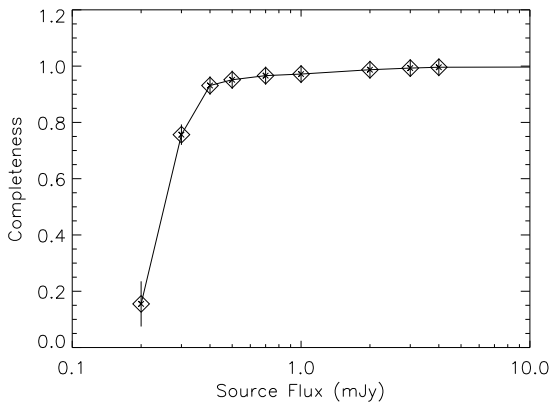,height=8cm,angle=90}
\caption{24$\mu$m Completeness function. 
Completeness as a function of flux density for the 24$\mu$m observations. Completeness is determined by injecting artificial sources into the map and then determining what fraction are recovered by our source detection technique. A thousand artificial sources are used for the determination of each point on the completeness function.}
\end{figure}

\subsection{24$\mu$m Counts}

\begin{figure*}
\label{counts24}
\epsfig{file=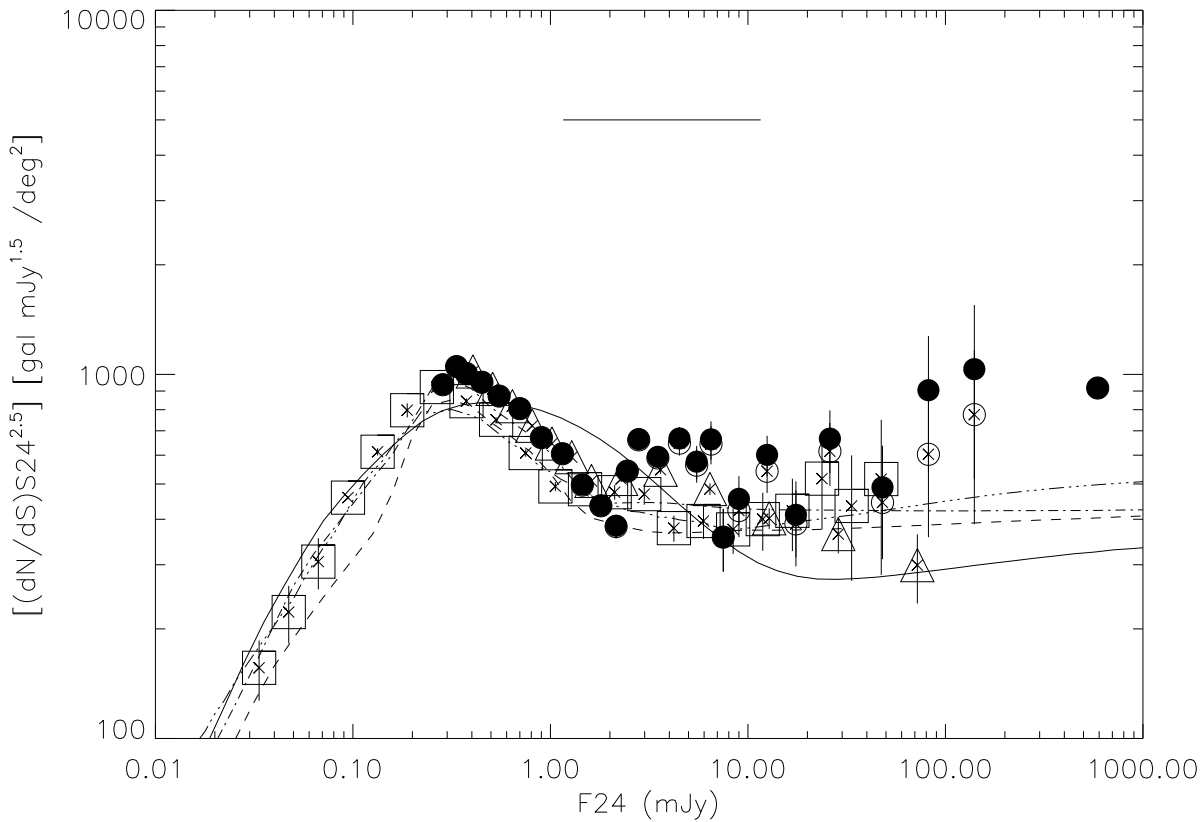,height=16cm,angle=90}
\caption{
Euclidean Normalized Differential Counts for our 24$\mu$m data compared to other counts in the literature and to theoretical models. Data from this work is shown as solid dots for raw counts and open circles for the counts where objects identified with cluster galaxies in DC 0428-43 have been removed. Literature counts are shown as follows: open squares: Papovich et al., (2004); open triangles: Shupe et al. (2008). The models are plotted as lines, including: solid line: Rowan-Robinson (2009) model; dashed line: Pearson et al. (2009) burst mode model; dash-dot line: Pearson et al. (2009) bright mode model; dash-three-dots line: Lagache et al (2004) model. The horizontal line above the counts indicates the range of flux densities expected for $0.2 L^*_{24} < L _{24}< 2L^*_{24}$ objects in the z=0.04 cluster in this field and for which some enhancement in counts might be expected (see section 6.3).}
\end{figure*}

The 24$\mu$m differential counts for the ADF-S field are shown in Fig. 6, alongside a variety of data (Papovich et al., (2004), Shupe et al., (2008)) and models from the literature (Lagache et al. (2004), Rowan-Robinson (2009)) as well as the Pearson number count models (Pearson \& Khan, 2009), described in somewhat more detail below. It should be noted that this is not a comprehensive list of available models. Many others are available, but we restrict the comparison to these models to compare the current 'baseline' model, from Lagache et al. (2004) to two more recent models based on SWIRE and other recent Spitzer observations while keeping the comparison diagrams simple enough for easy interpretation. The counts are also given in Table 1. Our data at 24$\mu$m largely covers flux density regions already covered by other surveys, but our greater area results in a larger number of sources and thus smaller Poisson error bars. Additional errors in the determination of the counts can come from large scale structures (LSSs) in the galaxy distribution. The cluster DC0428-43, which lies in our survey region and which is discussed more below, is a specific example of this effect, but generic LSS will also add to the uncertainty in the counts. We assess the uncertainties in the counts coming from LSS effects by extracting counts in a number of sub-regions in our survey field and performing a counts-in-cells analysis (see eg. Wall \& Jenkins, 2003). These are then added in quadrature to the Poisson errors. The LSS errors dominate the Poisson errors in all except the brightest bins. At the brightest flux densities, which are not well constrained by other surveys, we find a moderate excess of counts beyond what is expected on the basis of models. This, however, is a result of the foreground cluster at z=0.04 that covers $\sim$1/4 of this field (see below). In other respects, our 24$\mu$m counts are in good agreement with previous observational results.

\begin{table*}
\begin{tabular}{cccccccc}\hline
Upper (mJy)&Lower (mJy)&Av. Flux density (mJy)&Obs. Number&Completeness&Counts (gal sr$^{-1}$Jy$^{1.5})$&log[$(dN/dS)S^{2.5}$]\\ \hline
      1000.0 & 180.0   &   590.0  & 1       &1.00165&	918$\pm$918$\pm$0&2.96$^*$\\
      180.0   & 100.0     & 140.0   & 4      & 1.00345&	1034$\pm$516$\pm$0&	3.01$^*$\\
      100.0    &  64.0    &  82.0    & 6     & 1.00369&	905$\pm$369$\pm$0&	2.96$^*$\\
      64.0     &   32.0     & 48.0    &  11    &  1.00382&	489$\pm$148$\pm$391&	2.69$^*$\\
      32.0     &   20.0     & 26.0    &  26     & 1.00391&	666$\pm131$$\pm$266&		2.82$^*$\\
      20.0      &  15.0     & 17.5    &  18      &1.00395&	411$\pm$97$\pm$82&	2.61$^*$\\
      15.0      &  10.0     & 12.5    &  61     & 1.00397&	601$\pm$77$\pm$120&	2.78$^*$\\
      10.0     &      8.0     & 9.0      & 42      & 1.00398&	455$\pm$70$\pm$228&	2.66$^*$\\
      8.0      &7.0            & 7.5      & 26      & 1.00399&	357$\pm$70$\pm$193&	2.55$^*$\\
      7.0      &6.0            & 6.5      & 69      & 1.00399&	662$\pm$80$\pm$345&	2.82$^*$\\
      6.0      &5.0            & 5.5      & 91      & 1.00399&	576$\pm$60$\pm$207&	2.76$^*$\\
      5.0      &4.0           & 4.5       & 174    & 1.00400&	667$\pm$51$\pm$160&	2.82$^*$\\
      4.0      &3.0       &     3.5       & 290    & 1.00550&	593$\pm$35$\pm$83&	2.77$^*$\\
      3.0      &2.6         &   2.8       & 226    & 1.00820&	664$\pm$44$\pm$173&	2.82$^*$\\
      2.6      &2.3         &   2.45     & 193    & 1.01030&	542$\pm$40$\pm$141&	2.73$^*$\\
      2.3      &2.0         &   2.15     & 189    & 1.01210&	384$\pm$28$\pm$38&	2.58\\
      2.0      &1.6         &   1.80     & 444    & 1.01620&	436$\pm$21$\pm$44&	2.64\\
      1.6      &1.3         &   1.45     & 650    & 1.02180&	498$\pm$20$\pm$40&	2.70\\
      1.3      &1.0          &  1.15     & 1405  & 1.02660&	606$\pm$16$\pm$61&	2.78\\
      1.0     &0.8          &   0.90     & 1907  & 1.03100&	671$\pm$15$\pm$54&	2.82\\
     0.8     &0.6          &   0.70      & 4274  & 1.03500&	806$\pm$12$\pm$8&	2.91\\
     0.6     &0.5           &  0.55      & 4179  & 1.04700&	872$\pm$13$\pm$35&	2.94\\
     0.5     &0.40         &  0.45      & 7424  & 1.06250&	952$\pm$11$\pm$48&	2.98\\
     0.4     &0.35         &  0.375    & 5785  & 1.13600&	1006$\pm$13$\pm$30&	3.00\\
     0.35    & 0.32       &  0.335    & 4424  & 1.24000&	1051$\pm$16$\pm$84&	3.02\\
     0.32     &0.25       &  0.285    & 8140  &  2.09150&	937$\pm$10$\pm$56&	2.97\\
\end{tabular}
\caption{ADF-S 24$\mu$m Number Counts. 
Columns give upper, lower and average flux density for each bin, the observed number of sources in this flux density range, the completeness correction factor for this flux density, the Euclidean normalized corrected counts and the log of these normalized counts. $^*$ indicates flux range where the cluster galaxies are likely to have increased the counts significantly above the field value. Two errors are given for the counts. The first of these is the Poisson error, the second is the LSS error estimated through a counts-in-cells analysis. The LSS error is given as 0 where there are too few sources in the field to permit the counts-in-cells analysis. The two faintest data points correspond to the 80\% and 50\% completeness levels of the catalog respectively.}
\end{table*}

\section{70 Micron Source Extraction and Counts}

\subsection{Source Detection and Extraction}

Sources are detected in the 70$\mu$m map using the APEX image extraction tool developed by IPAC as part of MOPEX for use with MIPS-Ge data (Makovoz \& Marleau, 2005) with an approach similar to that used by Frayer et al. (2006a, 2006b, 2009). This uses an empirical point source response function (PRF) to extract source flux densities. Sources were extracted down to 4$\sigma$ significance in regions where the sky is observed by MIPS at least 10 times. We remove spurious low significance sources by cross-matching this 70$\mu$m source list to the 24$\mu$m source list discussed above using a 9" matching radius. Since the latter is more than 100 times deeper in terms of flux density we expect very few genuine 70$\mu$m sources to lack 24$\mu$m detections. This process excludes 404 out of 1702 initial detections, the majority of which are in regions of the map with poor 70$\mu$m coverage and thus enhanced noise. A further 95 sources are excluded because there is a small region where the 70$\mu$m and 24$\mu$m maps do not overlap and we thus do not have 24$\mu$m counterparts for matching. This decreases the effective area of the 70$\mu$m catalog from 11.5 to 10.5 sq. deg.

\subsection{Completeness Correction and Reliability Test}

We assess the completeness of the catalog as a function of flux density by injecting false sources at a given flux density into the map and then determining what fraction are recovered by our source extraction techniques. An empirical noiseless point spread function image is used to add each of the point sources to the input map. The empirical PSFs were created using archival {\it Spitzer} data of 3C 273,
3C 279, and BL Lac.  Data from individual AORs for these three
galaxies were processed into images using the MIPS DAT and then
combined so as to filter out any extended, asymmetric emission.  See
Young et al. (2009) for additional details.

We add 1000 artificial point sources in each of the flux density regimes where we test completeness, for a total of nine flux density bands and 9000 artificial sources. The resulting completeness as a function of source flux density is shown in Figure 7. Linear interpolation of this completeness function is then used to determine completeness correction factors in the calculation of the differential number counts. We find that our catalog is complete at the 80\% and 50\% levels at fluxes of 0.03 and 0.024 Jy respectively.

We test reliability by using a negative map that otherwise matches the noise characteristics of the real map. This is produced in the same way as that used to test the reliability of the 24$\mu$m catalog. We find that our source list is close to 100\% reliable over the full range of fluxes discussed here, with reliability falling from 100\% to 99\%, 97\% and 98\% for the three faintest flux bins. This reliability assessment does not include the cross matching to 24$\mu$m sources so is a conservative estimate of the reliability of our catalog. When we attempt to match the false, negative sources to genuine 24$\mu$m sources using the approach detailed above we find no matches at all. We thus conservatively conclude that our 70$\mu$m catalog is at least 97\% reliable throughout the flux range.

\subsection{Photometric Correction}

A correction factor of 1.15 was derived by Frayer et al. (2009) in their analysis of COSMOS counts to account for the emission outside the APEX PRFs. We use our injected artificial sources to test the correction factor needed in our data by examining the ratio of input flux density to recovered flux density. At bright flux density levels, comparable to the calibration star HD180711 (S70 = 447.4mJy, Gordon et al., 2007) used to check the Frayer et al. (2009) corrections, we measure the same correction factor of 1.15. At lower flux densities though, where the bulk of the 70$\mu$m counts lie, we find that a somewhat smaller correction factor is needed, as shown in Figure 8. We apply a linear interpolation of this correction function to our data in the calculation of the differential number counts. The overall absolute flux density calibration of MIPS at 70$\mu$m is uncertain at the 5\% level given that few of our sources are bright enough to be affected by nonlinearities in the detectors (Gordon et al., 2007).

\begin{figure}
\label{comp70}
\epsfig{file=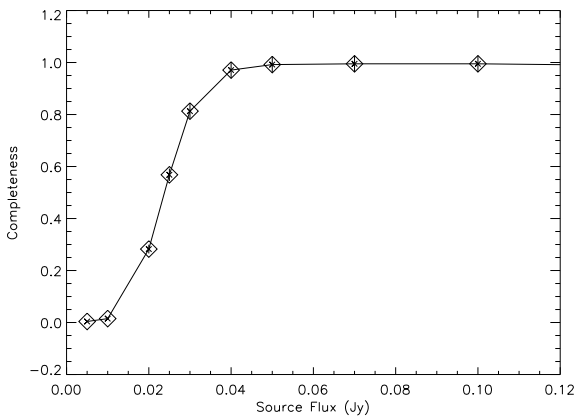,height=8cm,angle=90}
\caption{70$\mu$m Completeness function.
Completeness as a function of flux density for the 70$\mu$m observations. Completeness is determined by injecting artificial sources into the map and then determining what fraction are recovered by our source detection technique. A thousand artificial sources are used for the determination of each point on the completeness function.}
\end{figure}

\begin{figure}
\label{flux70}
\epsfig{file=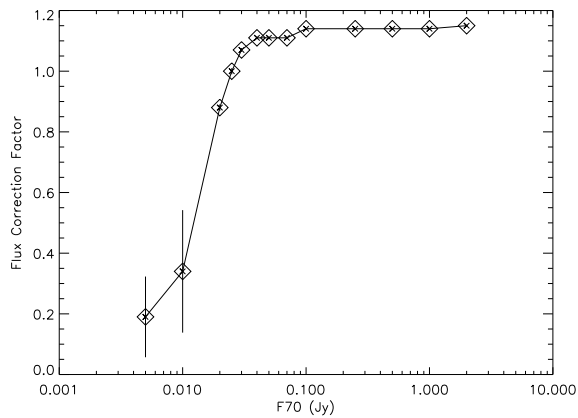,height=8cm,angle=90}
\caption{70$\mu$m Flux Density Correction function. 
Ratio of input flux density to output flux density for the artificial sources injected into the 70$\mu$m maps. This value is used to correct the photometry of the sources in our final catalog.}
\end{figure}

\begin{table*}
\begin{tabular}{cccccccc}\hline
Upper (Jy)&Lower (Jy)&Av. Flux density (Jy)&Obs. Number&Completeness&Counts (gal sr$^{-1}$Jy$^{1.5})$&log[$(dN/dS)S^{2.5}$]\\ \hline
      2.00    & 0.70       &1.35        &2.00000      &1.08050&      1104$\pm$      843 $\pm$1766&     3.04$^*$\\
     0.70      &0.40       &0.55        &9.00000      &1.04300&      2201$\pm$       765  $\pm$ 1321&    3.34$^*$\\
     0.40      &0.23       &0.315      &22.0000      &1.03260&      2334 $\pm$      514 $\pm$933&      3.37$^*$\\
     0.23      &0.14       &0.185      &29.0000      &1.01917&      1516 $\pm$      287  $\pm$ 303&    3.18$^*$\\
     0.14      &0.11       &0.125      &27.0000      &1.00917&      1573 $\pm$      306   $\pm$0&    3.20\\
     0.11      &0.09       &0.100      &36.0000      &1.00500&      1794$\pm$       300  $\pm$0&     3.25\\
    0.09       &0.075     &0.0825   &39.0000      &1.00500&      1602 $\pm$      258   $\pm$0&    3.20\\
    0.075     &0.062    &0.0685    &61.0000      &1.00522&      1816 $\pm$      234   $\pm$0&    3.26\\
    0.062     &0.050    &0.056      &95.0000      &1.00710&      1855 $\pm$      192  $\pm$ 0&    3.27\\
    0.050     &0.040    &0.045      &127.000      &1.01900&      1743 $\pm$      158  $\pm$ 0&    3.24\\
    0.040     &0.030    &0.035      &190.000      &1.13000&      1543 $\pm$      126  $\pm$0&    3.19\\
    0.030     &0.024    &0.027    &165.000      &1.54800&      1599 $\pm$      193   $\pm$0&    3.20\\
\end{tabular}
\caption{ADF-S 70$\mu$m Number Counts.
Columns give upper, lower and average flux density for each bin, the observed number of sources in this flux density range, the completeness correction factor for this flux density, the Euclidean normalized corrected counts and the log of these normalized counts. Errors on the counts are first the Poisson error and secondly the LSS error as determined by a counts-in-cells analysis. Where the LSS error is 0 the counts in cells analysis found no significant deviation in the counts in subfields beyond what would be expected by Poisson statistics. $^*$ indicates flux range where the cluster galaxies are likely to have increased the counts significantly above the field value. The two faintest data points correspond to the 80\% and 50\% completeness levels of the catalog respectively.}
\end{table*}
\subsection{70$\mu$m Counts}

The 70$\mu$m differential counts for the ADF-S are shown in Fig. 9 together with a variety of data and models from the literature. As with the 24$\mu$m counts this is not a comprehensive presentation of the full range of models currently available. The errors on the differential counts are a combination of Poisson counting errors and LSS noise calculated using counts-in-cells. Our data set new constraints on the counts at bright flux density levels and place tighter constraints at medium flux density levels than are currently available. They are in broad agreement with existing data where they overlap, though there is considerable scatter in all counts determinations at these wavelengths as a result of cosmic variance.  At bright flux density levels we see a moderate excess of sources over what is expected on the basis of theoretical models. This may be a result of sources in the foreground cluster at z=0.04 that lies in this field. This is also reflected in the large LSS errors found at bright flux levels.

\begin{figure*}
\label{counts70}
\epsfig{file=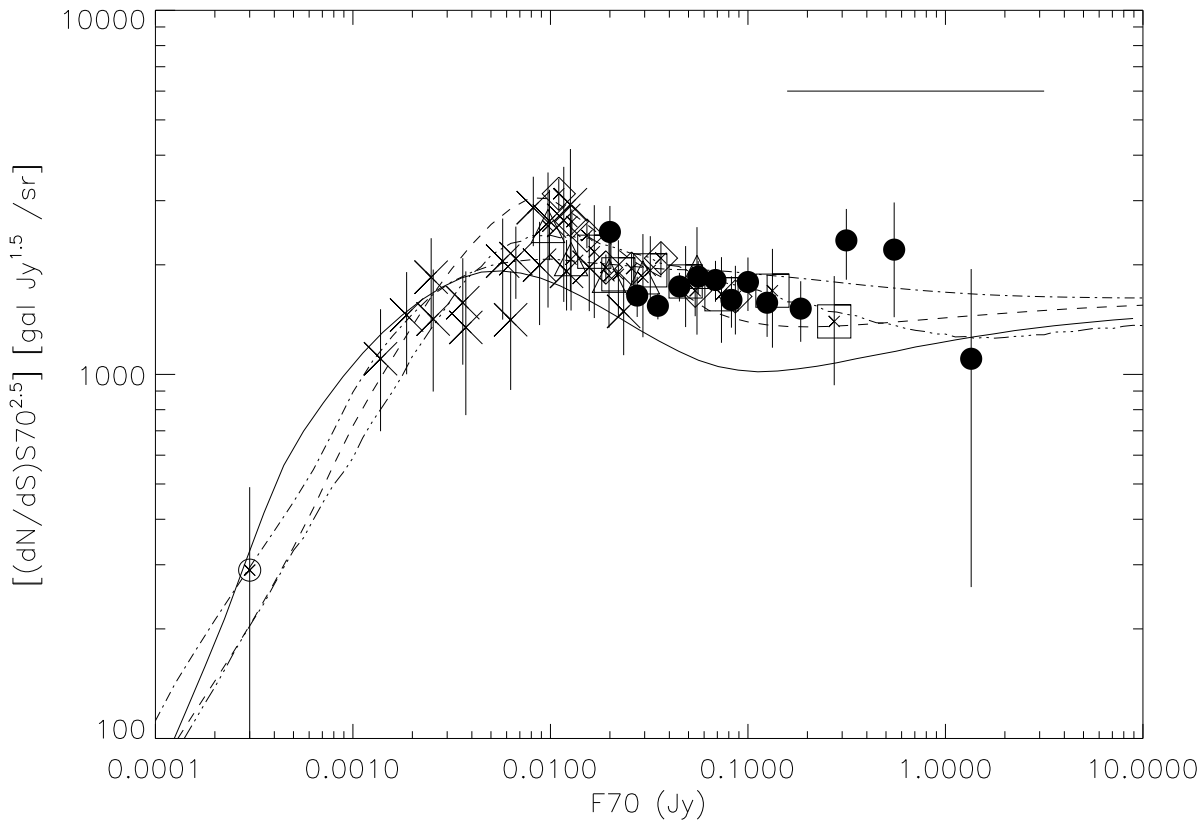,height=16cm,angle=90}
\caption{70$\mu$m Euclidean Normalized Differential Counts. 
Euclidean Normalized Differential Counts for our 70$\mu$m data compared to other counts in the literature and to theoretical models. Data from this work is shown as solid dots. Literature counts are shown as follows: open squares: Frayer et al. (2006a) FLS warm+main field; open diamonds: Frayer et al. (2009) COSMOS field; open triangles: Frayer et al. (2006a) FLS verification field; stars: Frayer et al. (2006b) GOODS-N field; open circle: confusion analysis point from Frayer et al. (2006b). The models are plotted as lines, including: solid line: Rowan-Robinson (2009) model; dashed line: Pearson (2009) burst mode model; dash-dot line: Pearson (2009) bright mode evolution model; dash-three-dots line: Lagache et al (2004) model. The horizontal line above the counts indicates the range of flux densities expected for $0.2 L^*_{70} < L _{70}< 2L^*_{70}$ objects in the z=0.04 cluster in this field and for which some enhancement in counts might be expected (see section 6.3).}
\end{figure*}

\section{Source Catalogs}

The source catalogs for the ADF-S produced from this {\it Spitzer} data are available via the electronic version of this paper. The supplied data includes position and flux densities for each source. Examples of the first lines of the catalog are given in tables 3 and 4 for the 24$\mu$m and 70$\mu$m catalogs respectively.

\begin{table*}
\begin{tabular}{cccccc}\hline
Source Number&RA&Dec&Positional Error (deg)&24$\mu$m Flux Density (mJy)&24$\mu$m Flux Density Error (mJy)\\ \hline
0&	66.65229&	-55.20967&0.0013& 	284.95&	0.067\\
1&	68.49929&	-55.04489&0.0017& 	271.58&	0.020\\
2&	67.20997&	-54.50196&0.0010&	188.45&	0.020\\
3&	70.87384&	-52.54578&0.0011&	173.64&	0.020\\
4&	68.49582&	-54.16836&0.0014&	141.12&	0.020\\
5&	71.86896&	-52.33193&0.0012&	115.68&	0.059\\
6&	72.73022&	-53.46116&0.0010&	115.21&	0.060\\
7&	69.89899&	-53.01719&0.0010&	110.85&	0.071\\
8&	69.90396&	-53.01198&0.0012&	109.52&	0.056\\
9&	74.13552&	-53.02721&0.0011&	101.69&	0.020\\
\end{tabular}
\caption{First ten lines of the 24$\mu$m {\it Spitzer} ADF-S catalog}
\end{table*}

\section{Discussion}

\subsection{24-70$\mu$m Colours}

Each of our 70$\mu$m sources is confirmed through cross identification at 24$\mu$m, so we can immediately examine the 70-to-24 $\mu$m colour distribution for our sources. This is shown in Fig. 10. The colour distribution is consistent with that found by Frayer et al. (2006b) for the {\it Spitzer} FLS sources. If we assume a power law for the SEDs of our objects between 24 and 70$\mu$m of the form $f_{\nu} \propto \nu^{-\alpha}$ we find our sources have an average spectral index of $\alpha$= 2.8$\pm$0.3 compared to Frayer et al.'s value of 2.4$\pm$0.4. In the absence of redshifts, we cannot examine the detailed physics behind the variation in colour. We note that sources with a log(70/24) $<$0.5 are likely to have their far-IR emission powered purely by an AGN (Frayer et al., 2006b). We find 18 sources in our catalog with log(70/24)$<$ 0.5 which can thus be regarded as candidate AGNs. However, one of these sources has such an exceptionally low ratio, log(70/24)  $<$-1, that it is consistent with a Rayleigh-Jeans stellar continuum. We identify this source as $\alpha$ Doradus, a bright binary star system (V=3.25 and V=4.3) whose primary is a chemically peculiar variable star (Heck et al., 1987). It should be noted that this source is successfully identified as a star through our K-[24] colour analysis and is thus not included in our consideration of the galaxy counts. We thus conclude that 17 of our sources have their mid-to-far-IR luminosity powered by an AGN. This fraction of AGN sources is consistent with the FLS results. Figure 10 also shows the range of colours that would be found for three archetypical objects over a range of redshifts where they might be detected. These include the local, quiescent spiral galaxy M100, shown from z=0 to z=1, a generic AGN SED, the local starburst M82, and the nearest ULIRG, Arp220, all shown from z=0--5. This suggests that many of the sources detected at both 70 and 24$\mu$m in this sample are actively star forming objects like M82 or Arp220.

\begin{figure}
\label{col70}
\epsfig{file=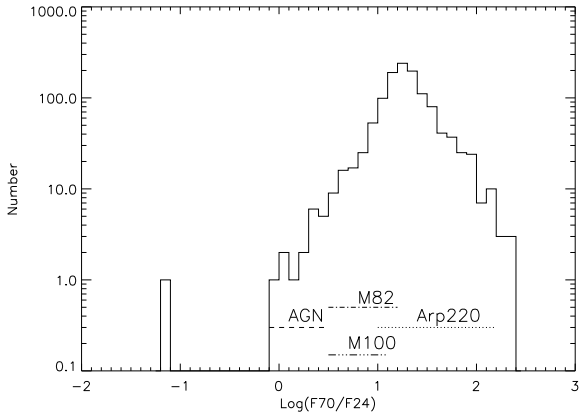,height=8cm,angle=90}
\caption{Histogram of the 70 to 24 $\mu$m flux density ratio for the sources in the 70$\mu$m catalog. The colour distribution is similar to that seen by Frayer et al. (2006b) in the FLS.  The single source with a very low 70/24 ratio is a bright binary star system, $\alpha$ Doradus whose flux density ratio is consistent with its 70 and 24 $\mu$m emission coming solely from the stellar photospheres. Also shown are the ranges of colour that would be found for various sources over a redshift range where they are likely to be detected. These include a generic AGN SED, the starburst M82 and the nearest ULIRG, Arp220, shown for z=0--5, and the lower luminosity quiescent local spiral galaxy M100 shown for z=0--1. Note that the colour ranges are not monotonic with redshift so, for example, the reddest colour for Arp220 occurs at z$\sim$1.5.}
\end{figure}

\begin{table*}
\begin{tabular}{ccccccc}\hline
Source Number&RA&Dec&Positional Error (deg)&70$\mu$m Flux Density (Jy)& Signal/Noise ratio\\ \hline
0.0&	66.96191&	-55.05889& 		0.00068&	0.92&	125.0\\
1.0&	68.49582&	-54.16836& 		0.0014&	0.91&	134.6\\
2.0&	68.81533&	-54.31517&		0.0009&	0.80&	122.7\\
3.0&	69.10008&	-54.41525&		0.0008&	0.68&	111.2\\
4.0&	68.43724&	-53.77945&		0.00087&	0.66&	95.2\\
5.0&	69.54994&	-54.37454&		0.00054&	0.60&	94.4\\
6.0&	68.59667&	-54.69228&		0.0013&	0.60&	96.6\\
7.0&	70.02633&	-54.34266&		0.00058&	0.59&	86.1\\
8.0&	73.56086&	-53.36012&		0.00069&	0.59&	83.7\\
9.0&	66.76700&	-54.26271&		0.00059&	0.56&	78.9\\
\end{tabular}
\caption{First ten lines of the 70$\mu$m {\it Spitzer} ADF-S catalog. The positions given in the 70$\mu$m catalog are those of the associated 24$\mu$m sources since the 24$\mu$m positions are more accurate than those obtained at 70$\mu$m.}
\end{table*}

\subsection{Count Models}

We compare our observed number counts, and those from the literature, to a number of different count models. These include models from Lagache et al. (2004) and Rowan-Robinson (2009) as well as two models from Pearson (Pearson \& Knan, 2009) which have yet to be fully described in the literature.

\subsubsection{Pearson Count Models}

The Pearson model incorporates  an infrared backward evolution framework following the models of Pearson (2005) and Pearson et al. (2007). These models were previously successfully used to reproduce the  combined mid-infrared source counts from {\it ISO} \& {\it {\it Spitzer}} at 15$\mu$m \& 24$\mu$m and have recently been successfully applied to the source counts at submillimetre wavelengths (Pearson \& Khan, 2009)  
The model uses the  60$\mu$m luminosity function derived from the {\it IRAS} Point Source Catalogue (Saunders et al., 2000) to represent cool and warm galaxy populations defined by local  {\it IRAS} colours, where cool 100$\mu$m/60$\mu$m cirrus-like colours represent the quiescent, non-evolving normal galaxy population and warmer 100$\mu$m/60$\mu$m colours represent evolving star-forming galaxies. To model the AGN, the luminosity function derived from the 15$\mu$m selected sample of Matute et al. (2006) is used. The model incorporates a suite of spectral energy distributions to model the cool quiescent, warm star-forming and AGN populations. The spectral templates for the quiescent normal galaxy population have been selected from the libraries described in Efstathiou $\&$ Rowan-Robinson (2003). The adopted star-forming templates are selected as a function of increasing luminosity for starburst ($L_{IR}<10^{11}L_{\sun}$), LIRG ($\rm 10^{11}L_{\sun} < L_{IR}< 10^{12}L_{\sun}$) and ULIRG ($ L_{IR}> 10^{12}L_{\sun}$) populations. A total of 7 SEDs (2 starburst, 3 LIRG, 2 ULIRG) are selected from the starburst model template libraries of  Efstathiou et al. (2000) to ensure a reasonable variation in the mid-infrared PAH features in an attempt to avoid features in the source counts caused by a single choice of template for all sources. The AGN template is taken from the  tapered disc dust torus models of Efstathiou \& Rowan-Robinson  (1995). All the spectral templates have been shown to provide good fits to  IRAS, ISO and {\it Spitzer} selected samples of infrared populations (Efstathiou et al. (2000) , Rowan-Robinson et al.(2004), Rowan-Robinson et al.  (2005)).

The general Pearson Model framework is available in two evolutionary flavours referred to as the  $bright$ and $burst$ models  categorized by their dominant populations of  starburst (M82 like) and LIRG/ULIG (Arp 220 like) sources respectively. The  {\it bright} model broadly follows the evolutionary  scenario of Pearson \& Rowan-Robinson (1996) and assumes all active populations  evolve in both luminosity and density as a double power law, of evolutionary strength $k$ of the form $(1+z)^{k}$. The evolution rises steeply to a redshift of unity and more shallowly thereafter. The {\it burst} model was originally introduced in Pearson (2001) and assumes that the more luminous  LIRG \& ULIRG populations evolve exponentially (of the form  $1 + k  . exp[-{(z-z_{p})^{2}\over{2\sigma ^2}}]$, where $k$ and $\sigma$ are evolutionary parameters) from the present epoch to some peak redshift $z_{p}$ ensuring both local rarity and the emergence of the population towards a peak redshift $\sim$1. The $bright$ and $burst$ models are explained in detail in Pearson (2009).

\subsubsection{Comparison to Observed Counts}

At 24$\mu$m all of the models perform reasonably well with no gross disagreements with the observations. However,  none of the models are an ideal fit to the data. The Pearson burst model underpredicts the counts below $\sim$0.3mJy while the other three models do well in this flux density regime. Between 0.3 and $\sim$3mJy all models except the Rowan-Robinson model match the data well, with the Rowan-Robinson model overpredicting the counts around 1-2mJy.  Counts at flux densities brighter than a few mJy are still quite poorly constrained, especially considering the problem with the cluster in the current data set. The best constraints come from SWIRE and suggest that all models except that of Rowan-Robinson are overpredicting the brightest counts. The Lagache model has the most significant disagreement in this respect. The all sky WISE survey  (Eisenhardt et al., 2009) will settle the issue of the brighter counts now that this satellite has been launched. This visual inspection of the counts marginally favours the Pearson bright model, with the Lagache model second followed by the Rowan-Robinson and Pearson burst model.

At 70$\mu$m, possibly because this is a wavelength regime that was not well studied prior to {\it Spitzer}, there are rather larger differences among the models and between the data and the models. The Rowan-Robinson model significantly underpredicts counts in the 0.02 to 1 Jy range and fails to place the peak of the normalized counts at the right flux denstiy. The Pearson burst model overpredicts counts in the 0.002 to 0.01 Jy range, predicting too broad a peak for the counts. The Pearson bright and Lagache models both visually match the data reasonably well.

We perform a quantitative test of the models against the data by calculating the sum of the $\chi^2$ values of the models compared to the data sets shown in Figures 6 and 9. These support the conclusions drawn from the visual inspections of the data. At 24$\mu$m the Pearson bright model has the lowest $\chi^2$ by a considerable margin, with less than half the $\chi^2$ value of the Rowan-Robinson model and less than a third the $\chi^2$ of the Lagache or Pearson burst models. At 70$\mu$m the results are less clear cut, with the Lagache and Pearson burst and bright models having very similar $\chi^2$ values while the Rowan-Robinson model has roughly three times this value.

We conclude that the Pearson bright model fits the counts at 24 and 70$\mu$m more accurately than the
other models examined here. Additional data at the bright
end of the 24 $\mu$m counts, at the fainter end of the 70$\mu$m
counts, and flux densities close to the peak of the 70$\mu$m counts
(i.e. at 0.01 Jy and fainter) would be the most useful for
constraining the models further.
There is currently only one {\em Spitzer} data set providing constraints to the 70$\mu$m counts at 0.01Jy and fainter, the GOODS-N counts from Frayer et al. (2006b), so more data in this regime are particularly needed. Such observations should be forthcoming from both the {\it Spitzer} Far-Infrared Deep Extragalactic Legacy (FIDEL) survey (Dickinson et al., 2007) and from observations with the PACS instrument on the Herschel Space Observatory. The latter is also able to probe deeper in small fields because of the fainter confusion limit provided by the 3.5m Herschel primary mirror, as demonstrated by initial results at 100 and 160$\mu$m from the PEP survey (Berta et al., 2010).

\subsection{The Local Cluster - DC 0428-43}

As discussed in the introduction, the ADF-S encompasses the z=0.04 galaxy cluster DC 0428-53 (Dressler, 1980; also known as Abell S0463). We might thus expect some enhancement in the counts seen at flux densities appropriate to $\sim L^*$ galaxies at the cluster redshift. We take estimates for the field galaxy population L$^*$ at 24$\mu$m from Babbedge et al. (2006) and for 70$\mu$m from Patel et al. (in preparation), yielding flux densities of 5.8mJy and 310mJy respectively, and show an indicative range of effect from 0.2---2 L$^*$ in Figures 3 and 6. As can be seen there is some indication of enhanced counts at the brighter end of this flux density range at both 24 and 70$\mu$m with less of an effect at fainter flux densities. This is to be expected as the contribution from cluster objects will be less significant at fainter flux densities where we detect greater numbers of objects in the field population. For example, in the 3.5mJy flux density bin at 24$\mu$m we detect 290 objects while Dressler et al. (1980) detects a total of only 131 optically selected galaxies in the cluster with 84 spectroscopically confirmed (Dressler \& Schectman, 1988). There is some indication of excess counts above model expectations or SWIRE observations in the brightest 24$\mu$m bins, equivalent to 10---20 L$^*_{24}$ at the cluster redshift. Cluster objects might be responsible for this if their 24$\mu$m luminosity is enhanced beyond that of the field population, possibly through enhanced star formation rates resulting from interactions between galaxies in different cluster components, and we do indeed find that four of our five brightest objects lie in the third of the field adjacent to the cluster. Such effects have been suggested elsewhere (Fadda et al., 2000). In this context it is interesting to note that Dressler \& Schectman (1988) find evidence for substructure in DC 0428-53, with a colder subsystem apparently merging with the main cluster. 

We attempt to remove the influence of the cluster from our counts by excluding from consideration those sources in the 24 and 70$\mu$m catalogs that are coincident with sources in the Dressler \& Schectman (1988) catalog of objects spectroscopically identified as cluster members. This is shown in Figs. 6 and 9. As can be seen this removes the enhancement at bright fluxes seen at 70$\mu$m, but does not have the same effect at 24$\mu$m. This would suggest that there are cluster members detected at 24$\mu$m that are not in the list of spectroscopically confirmed cluster members. Given that the 24$\mu$m observations are quite deep and that the optical spectroscopy is quite shallow, V$<$16, this is not unexpected. Further examination of this cluster and its galaxies is underway but is beyond the scope of this paper.

\section{Conclusions}

We present the results of 24 and 70$\mu$m {\it Spitzer} observations of the ADF-S region, which covers a contiguous field of $\sim$12 sq. deg.. We derive galaxy number counts from this data. They confirm that the galaxy population in these bands evolves strongly. We compare our own and literature counts to galaxy count models from Lagache et al. (2004), Rowan-Robinson et al. (2009) and two models by Pearson (Pearson \& Khan, 2009). We find that the Pearson bright model provides a reasonable fit to the counts in both bands, performing as well as, or better than, the more established Lagache and Rowan-Robinson models, but the Pearson burst model does not perform as well. None of the three best performing models provide a fully successful match to the data. We also discuss the contribution of galaxies within a local cluster, DC0428-43 at z=0.04 which lies within the ADF-S, to the counts and find indications that some cluster galaxies may have their 24$\mu$m luminosities enhanced beyond what is found in the field population. The full source catalogs derived from these observations are available in electronic form for download.
~\\~\\
{\bf Acknowledgements}
This work is based on observations made with the {\it Spitzer} Space Telescope, which is operated by the Jet Propulsion Laboratory, California Institute of Technology under a contract with NASA.This publication also makes use of data products from the Two Micron All Sky Survey, which is a joint project of the University of Massachusetts and the Infrared Processing and Analysis Center/California Institute of Technology, funded by the National Aeronautics and Space Administration and the National Science Foundation. This work has been funded, in part, by STFC. We thank M. Rowan-Robinson and G. Lagache for the provision of their count models and for useful discussions, Harsit Patel for cross-matching cluster galaxies, and the anonymous referee for their extensive and helpful comments which have signficantly improved the paper. 
\\~\\

\begin{appendix}
\section{Stellar Counts at 24 microns}

In Figure A1 we provide a plot of the number counts of sources identified as stars by our K-band vs. 24$\mu$m colour analysis.

\begin{figure}
\label{stars24}
\epsfig{file=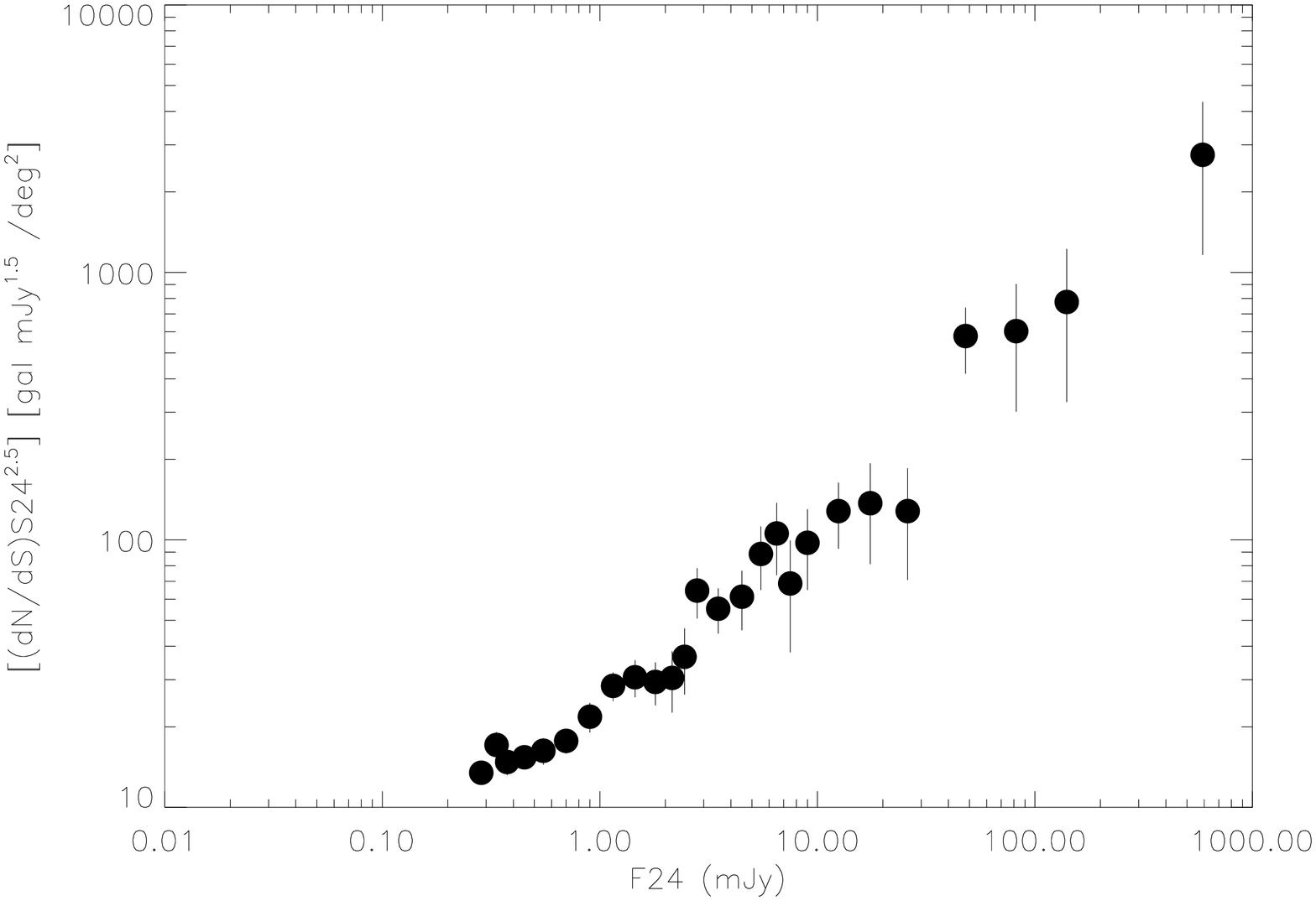,height=8cm,angle=90}
\caption{24$\mu$m Number Counts for objects identified as stars.}
\end{figure}

\end{appendix}

\end{document}